\documentstyle[12pt]{article}

\def\be{\begin{equation}}
\def\ee{\end{equation}}
\def\bea{\begin{eqnarray}}
\def\eea{\end{eqnarray}}

\begin{document}

\begin{flushright}
hep-th/0404209
\end{flushright}

\pagestyle{plain}

\def\e{{\rm e}}
\def\haf{{\frac{1}{2}}}
\def\tr{{\rm Tr\;}}
\def\goes{\rightarrow}
\def\ie{{\it i.e.}, }
\def\lra{\longrightarrow}
\def\Lra{\Longrightarrow}
\def\ol{\overline}

\begin{center}
\vspace{3cm} {\Large {\bf On The Classical Dynamics Of Charges In Noncommutative QED}}

\vspace{1cm}

Amir H. Fatollahi \footnote{Address after March 2004: Department of Physics, Alzahra University,
Tehran, 19938-91167, Iran} \hspace{3mm} and \hspace{3mm} Hossein
Mohammadzadeh

\vspace{.5cm}

{\it Institute for Advanced Studies in Basic Sciences (IASBS),\\
P. O. Box 45195, Zanjan 159, Iran}

\vspace{.3cm}

{\sl fatho@mail.cern.ch\\
mohammad@iasbs.ac.ir}

\vskip .5 cm
\end{center}

\begin{abstract}
One approach for formulating the classical dynamics of charged particles in
non-Abelian gauge theories is due to Wong. Following Wong's approach, we
derive the classical equations of motion of a charged particle in U(1) gauge
theory on noncommutative space, the so-called noncommutative QED. In the
present use of the procedure, it is observed that the definition of
mechanical momenta should be modified. The derived equations of motion manifest
the previous statement about the dipole behavior of charges in noncommutative
space.
\end{abstract}

\vspace{1.0cm}

\newpage

In the last years much attention has been paid to the formulation
and study of field theories on noncommutative
spaces. Apart from the abstract mathematical interests,
there are various physical motivations for doing so.
One of the original motivations has been to get ``finite"
field theories via the intrinsic regularizations which are
encoded in some of noncommutative spaces \cite{snyder}.
The other motivation comes from the unification aspects
of theories on noncommutative spaces. These unification
aspects have been the result of the ``algebraization" of
``space, geometry and their symmetries" via the approach
of noncommutative geometry \cite{connes}. Interpreting
the Higgs fields of the theories with spontaneously broken
symmetries as gauge fields in the discrete directions
of multi-sheet spaces is an example of this point of view on noncommutative
spaces \cite{connlott}. The other motivation refers to the natural appearance
of noncommutative spaces in some areas of physics, and the recent
one in string theory. It has been understood that string
theory is involved by some kinds of noncommutativities;
two examples are:
(1) the coordinates of bound states of $N$ D-branes are
represented by $N \times N$ Hermitian matrices \cite{9510135}, and
(2) the longitudinal directions of D-branes in the presence
of a B-field background appear to be noncommutative, as seen by the
ends of open strings \cite{9908142,CDS,jab-1}. In the latter case, we encounter
the spacetime in which the coordinates satisfy the canonical
commutation relation
\bea
{[}\hat{x}^\mu,\hat{x}^\nu{]}=i\theta^{\mu\nu},
\eea
in which $\theta^{\mu\nu}$ is a constant second rank tensor.
Since the coordinates do not commute, any
definition of functions or fields should be performed under a prescription
for ordering of coordinates, and a natural choice can be the symmetric one,
the so-called Weyl ordering. To any function $f(x)$ on ordinary space, one can assign
an operator $\hat{O}_f$ by
\bea
\hat{O}_f(\hat{x}):=\frac{1}{(2\pi)^n} \int  d^nk \; \tilde{f}(k)\; \e^{-ik\cdot \hat{x}}
\eea
in which $\tilde{f}(k)$ is the Fourier transform of $f(x)$ defined by
\bea
\tilde{f}(k)=\int d^nx \; f(x)\; \e^{ik\cdot x}.
\eea
Due to presence of the phase $\e^{-ik\cdot \hat{x}}$ in definition of
$\hat{O}_f$, we recover the Weyl prescription for the coordinates. In a revers
way we also can assign to any symmetrized operator a function or field
living on the noncommutative plane. Also, we can assign to product of any
two operators $\hat{O}_f$ and $\hat{O}_g$ another operator as
\bea
\hat{O}_f\cdot \hat{O}_g =: \hat{O}_{f\star g}
\eea
in which $f$ and $g$ are multiplied under the so-called $\star$-product defined by
\bea
(f\star g)(x)=\e^{\frac{i}{2}\theta^{\mu\nu}\frac{\partial}{\partial x^\mu}\frac{\partial}{\partial y^\nu}}f(x)g(y)\mid_{y=x}
\eea
By these all one learns how to define physical theories in noncommutative spacetime,
and eventually it appears that the noncommutative field theories are defined
by actions that are essentially the same as in ordinary spacetime, with the exception that
the products between fields are replaced by $\star$-products; see \cite{reviewnc} as review.
Though $\star$-product itself is not commutative (\ie $f\star g \neq g\star f$)
the following identities make some of calculations easier in field theories:
\bea
&&\int(f\star g)(x)d^{n}x=\int(g\star f)d^{n}x=\int (f\cdot g)(x)d^{n}(x)\\
&&\int(f\star g\star h)(x)d^{n}x=\int(f\cdot (g\star h))(x)d^{n}x=\int((f\star g)\cdot
h)(x)d^{n}x\\
&&\int(f\star g\star h)(x)d^{n}x=\int(h\star f\star g)(x)d^{n}x=\int(g\star h\star f)(x)d^{n}x
\eea
By the first two ones we see that, in integrands always one of the $\star$'s can be removed.

The noncommutative QED (NCQED) is given by the action
\bea
S=\int d^{4}x\bigg(-\frac{1}{4}F_{\mu\nu}F^{\mu\nu}
-\overline{\psi}\gamma^{\mu}(\partial_\mu -ig A_{\mu}\star)\psi-\frac{imc}{\hbar}\overline\psi\psi\bigg)
\eea
in which field strength is defined by
\bea
F_{\mu\nu}=\partial_{\mu}A_{\nu}-\partial_{\nu}A_{\mu}-ig[A_{\mu},A_{\nu}]_{MB}
\eea
in which the commutator is that by Moyal, defined by
\bea
[A_{\mu},A_{\nu}]_{MB}=A_{\mu}\star A_{\nu}-A_{\nu}\star A_{\mu}
\eea
The action is invariant under the transformations
\bea
\psi&\lra&\psi^{\prime}=U\star\psi\nonumber\\
\overline{\psi}&\lra&\overline{\psi}^{\prime}=\overline{\psi}\star U^{-1}\nonumber\\
A_{\mu}&\lra&{A^{\prime}}_{\mu}=U\star A_{\mu}\star U^{-1}+\frac{i}{g}\;U\star\partial_{\mu}U^{-1}
\eea
in which $U(x)$ is the $\star$-phase ($U\star U^{-1}=U^{-1}\star U=1$)
defined by a function $\lambda(x)$ via the $\star$-exponential:
\bea
U(x)=\exp_{\star}(i\lambda)=1+i\lambda-\frac{1}{2}\lambda\star\lambda+\cdots
\eea
Under the gauge transformation, the field strength transforms as
\bea
F_{\mu\nu}&\lra& F^{\prime}_{\mu\nu}=U\star F_{\mu\nu}\star U^{-1}
\eea
We mention that the transformations of gauge fields as well as the field strength
look like to those of non-Abelian gauge theories. Besides, we see that the
pure gauge field sector of action contains terms which are responsible for interaction
between the gauge particles, again as the situation we have in non-Abelian gauge
theories.

Among others, there is one approach due to Wong \cite{wong} for derivation of the
classical equations of motion of particles that have non-Abelian charges. In this formulation
there are a couple of equations among which one is for the dynamics of the charged particle in
spacetime, and one for the dynamics of isospin charge of particle, as an internal degrees
of freedom. The former is analogous to Lorentz force in electro-magnetism.
Noting the non-Abelian nature of NCQED, it is quite reasonable to
use the approach by Wong to derive the classical equations of motion for charges in
NCQED.

Let us review shortly Wong's approach in the next lines; see \cite{book}
\footnote{Note missing $i=\sqrt{-1}$ in before of mass term in \cite{book}.}.
The equations of motion for the fermionic matter field in fundamental representation
and in the presence of background field $A_\mu(x)$ is:
\bea
\gamma^\mu (\partial_\mu - i g A_\mu^a \hat{T}_a) \Psi(x) + \frac{imc}{\hbar} \Psi(x)=0
\eea
in which $\hat{T}_a$'s are the generators of group, satisfying $[\hat{T}_a,\hat{T}_b]=if_{ab}^c \hat{T}_c$.
Viewing this equation as a Schrodinger equation (recalling $\partial_0=c^{-1}\partial_t$)
one reads the Hamiltonian as
\bea
\hat{H}=c\alpha^i(\hat{p}_i-g\hbar A_i^a(\hat{x}) \hat{T}_a)+mc^2 \beta - gc\hbar A_0^a(\hat{x})\hat{T}_a
\eea
in which $\alpha^i$ and $\beta$ are the Dirac matrices, and $\hat{p}_i$ is sitting
for $-i\hbar\partial_i$. In the Heisenberg picture, one obtains
the equations of motion for operators as:
\bea
\dot{\hat{x}^i}&=&\frac{i}{\hbar} [\hat{H}, \hat{x}^i]= c\alpha_i,\\
\dot{\hat{p}}_i&=&\frac{i}{\hbar} [\hat{H}, \hat{p}_i]=
gc\hbar (\alpha^j\partial_iA_j^a + \partial_i A_0^a) \hat{T}_a,\\
\dot{\hat{T}}_a&=&\frac{i}{\hbar} [\hat{H}, \hat{T}_a]= -g f_{ab}^c (\dot{\hat{x}}_i A_i^b + c A_0^b) \hat{T}_c
\eea
By defining the mechanical momenta by $\hat{\pi}_i:= \hat{p}_i -g\hbar A_i(\hat{x})$,
one then gets
\bea
\dot{\hat{\pi}}_i=g \hbar (c F_{i0}^a + \dot{\hat{x}}^j F_{ij}^a ) \hat{T}_a
\eea
in which the $F_{\mu\nu}$'s are the field strength of the non-Abelian gauge theory,
defined by $F_{\mu\nu}^a=\partial_\mu A_\nu^a - \partial_\nu A_\mu^a -gf^a_{bc} A_\mu^b A_\nu^c$.
By comparison this equation with that of electro-magnetism Wong suggests the following for the
non-Abelian case:
\bea
m\ddot{\xi}_\mu=g (F_{\mu\nu}^a T_a) \dot{\xi}^\nu
\eea
in which $\xi^\mu(\tau)$ represents the world-line of the particle, and we used
the change $\hbar\hat{T}_a \rightarrow T_a$. The dot in the equation is for derivative
with respect to the proper-time. We mention that in the above equation $T_a$'s can not be and
are not supposed to be operators (\ie matrices) anymore, while we interpret them as number functions
capturing the degrees of freedom coming from group structure, satisfying the equations of motion:
\bea
\dot{T}_a + g  \dot{\xi}^\mu f_{ab}^c  A_\mu^b T_c = 0.
\eea
From this we learn that the group degrees of freedom, also known as isotopic spin, performs a
precessional motion: $d/d\tau (T_a T^a)=0$.

Now we use Wong's method for the case of NCQED, and we do this in the first order of
noncommutativity parameter $\theta_{\mu\nu}$. The Lagrangian in this order is:
\bea
L=-\frac{1}{4}F^{\mu\nu}F_{\mu\nu}-\ol{\psi}\gamma^{\mu}(\partial_{\mu}-igA_{\mu})\psi
-\frac{1}{2}g\ol{\psi}\gamma^{\mu}\theta^{\alpha\beta}\partial_{\alpha}A_{\mu}\partial_{\beta}\psi-\frac{imc}{\hbar}\ol{\psi}\psi + O(\theta^2)
\eea
in which the field strength is
\bea
F_{\mu\nu}=\partial_{\mu}A_{\nu}-\partial_{\nu}A_{\mu}+g\theta^{\alpha\beta}\partial_{\alpha}A_{\mu}\partial_{\beta}A_{\nu}
+O(\theta^2).
\eea

The action corresponding to the Lagrangian (23) is invariant under the
first-order transformations in $\theta$
\footnote{The first-order transformations can be obtained by noting the fact that the $\star$-power of a function $f(x)$
behaves: $f_\star^n:=f\star f \star \cdots \star f = f^n+O(\theta^2)$, and hence we have
$U(x)=\exp_{\star}(i\lambda)= \e^{i\lambda}+O(\theta^2)$.}:
\bea
\psi&\lra&\psi^{\prime}= \e^{i\lambda} (\psi - \frac{\theta^{\mu\nu}}{2} \partial_\mu\lambda \partial_\nu\psi)+O(\theta^2)\nonumber\\
\overline{\psi}&\lra&\overline{\psi}^{\prime}=\e^{-i\lambda} (\overline{\psi} - \frac{\theta^{\mu\nu}}{2} \partial_\mu\lambda
\partial_\nu\overline{\psi})+O(\theta^2)\nonumber\\
A_{\mu}&\lra&{A^{\prime}}_{\mu}= A_\mu+\frac{1}{g} \partial_\mu \lambda - \theta^{\alpha\beta} \partial_\alpha \lambda\;
\partial_\beta A_\mu - \frac{\theta^{\alpha\beta}}{2g} \partial_\alpha\lambda \;\partial_\beta\partial_\mu \lambda +O(\theta^2)
\eea

The equation of motion for $\psi$ is obtained to be:
\bea
\gamma^{0}(\partial_{0}-igA_{0})\psi+\gamma^{i}(\partial_{i}-igA_{i})\psi+\frac{1}{2}g\gamma^{\mu}\theta^{\alpha\beta}
\partial_{\alpha}A_{\mu}\partial_{\beta}\psi+\frac{imc}{\hbar}\psi=0
\eea
Here after we assume that noncommutativity is just for
spatial directions: $\theta^{0\mu}=\theta^{\mu 0}=0$. So, the above
equation appears in the form:
\bea
\gamma^{0}(\partial_{0}-igA_{0})\psi+\gamma^{i}(\partial_{i}-igA_{i})\psi+\frac{1}{2}g\theta^{ij}\gamma^{\mu}\partial_{i}A_{\mu}\partial_{j}\psi
+\frac{imc}{\hbar}\psi=0 \eea Again viewing this as a Schrodinger
equation we read the corresponding Hamiltonian as \bea
\hat{H}=-gc\hbar A_{0}+c\alpha^{i}(\hat{p}_i-g\hbar
A_i)+\frac{1}{2}gc\theta^{ij}\alpha^\mu\partial_iA_\mu\hat{p}_j+mc^{2}\beta
\eea
in which we have used
$\alpha^{\mu}=(\alpha^0,\alpha^{k})=(I,\alpha^{k})$. The
Heisenberg equations of motion are derived for the operators as
well:
\bea
\dot{\hat{x}^l}&=&\frac{i}{\hbar}[\hat{H},\hat{x}^l]=c\alpha^{l}+\frac{1}{2}gc\theta^{il}\alpha^{\mu}\partial_{i}A_{\mu}\\
\dot{\hat{p}_l}&=&\frac{i}{\hbar}[\hat{H},\hat{p}_{l}]=
gc\hbar\alpha^{\mu}\partial_{l}A_{\mu}-\frac{1}{2}gc\theta^{ij}\alpha^{\mu}\partial_{l}\partial_{i}A_{\mu}\hat{p}_j
\eea
From the first equation we have  $c\alpha^l=\dot{\hat{x}^l}-\frac{1}{2}gc\theta^{il}\partial_{i}A_{0}
-\frac{1}{2}g\theta^{il}\dot{\;\hat{x}^k}\partial_{i}A_{k}+O(\theta^2)$; this will be used for
later replacements. For the case of NCQED, we see that the interaction between fermions and gauge
fields is different in comparison with ordinary (Abelian and non-Abelian) gauge theories.
For the present case we have the following as mechanical momenta
\bea
\hat{\pi}_l=\hat{p_{l}}-g\hbar A_{l}(\hat{x}) +\frac{1}{2} g\theta^{ij}
\partial_iA_l(\hat{x}) \hat{p}_j
\eea
This form of mechanical momenta can be read also from the covariant derivative of
NCQED, $D_{\mu}\psi=\partial_{\mu}\psi-igA_{\mu}\star\psi$, which
changes by similarity transformation under gauge transformations.
After this modification to the Wong's approach, one can calculate
the time derivative as
\bea
\dot{\hat{\pi}_l}=\dot{\hat{p}_l}-g\hbar\partial_t A_l-ig[\hat{H},A_{l}]
+\frac{1}{2} g \theta^{ij} \partial_t\partial_i A_l \hat{p}_j
+\frac{1}{2} g \theta^{ij} \frac{i}{\hbar} [\hat{H}, \partial_iA_l \hat{p}_j]
\eea
After sufficient manipulations and replacements, and omitting hats we obtain
\bea
\dot{\pi}_l&=& gc\hbar(\partial_lA_0-\partial_0A_l) +g\hbar \dot{x}^i (\partial_lA_i-\partial_i A_l)\nonumber\\
&+&\frac{1}{2}gc\theta^{ij} p_j (\partial_0\partial_i A_l - \partial_l \partial_i A_0)
+\frac{1}{2}g \theta^{ij} \dot{x}^k p_j (\partial_k\partial_i A_l - \partial_l\partial_i A_k)\nonumber\\
&-&\frac{1}{2} g^2 c \hbar \theta^{ij} (\partial_i A_0 +\frac{1}{c} \dot{x}^k \partial_i A_k)
(\partial_j A_l + \partial_l A_j )+O(\theta^2)
\eea
By defining the field strengths:
\bea
&&f_{\mu\nu}=\partial_\mu A_\nu-\partial_\nu A_\mu\\
&&F_{\mu\nu}=f_{\mu\nu}+g\theta^{ij}\partial_i A_{\mu}\partial_j A_{\nu}
\eea
and by setting $g\hbar c=e$ and $\mu^i:=\frac{1}{2}gc \theta^{ij} p_j= \frac{e}{2\hbar}\theta^{ij} p_j$,
we get
\bea
\dot{\pi}_l&=& e (F_{l0} +\frac{1}{c}  \dot{x}^i F_{li})\\
&+&\mu^i \partial_i f_{0l} +\frac{1}{c} \mu^i  \dot{x}^k \partial_i f_{kl}
+ \frac{1}{c} \dot{\mu}^i f_{li}+O(\theta^2)
\eea
The first two terms are easily understood as the dynamics of a charged particle in the
background of noncommutative field strength $F_{\mu\nu}$. To understand the other
terms, we compare the result with those of a dipole electric in the background of
ordinary electro-magnetic fields. The corresponding Lagrangian for a point-like electric
dipole can be easily derived by considering the dynamics of two equal mass particles with
opposite charges $q$ and $-q$, while their relative distance $\ell$ is small,
defining the electric dipole $\vec{\mu}:= q \vec{\ell}$. So the starting point is
\bea
L=\frac{1}{2} m \dot{\vec{x}}^2 +\frac{1}{c} \mu^i \dot{x}^i f_{ji} + \mu^i f_{i0}
\eea
So the equation of motion for the position of dipole appears as
\bea
m\ddot{x}_k=\mu^j \partial_k f_{0j} -\frac{1}{c} \dot{\mu}^j f_{jk}
+\frac{1}{c} \mu^j \dot{x}^i (\partial_k f_{ji} - \partial_i f_{jk}).
\eea
After using the Maxwell
equation of ordinary electro-magnetism \footnote{And using  the identity
$\epsilon_{jil}\partial_k - \epsilon_{jkl}\partial_i = \epsilon_{kil}\partial_j
- \epsilon_{kij}\partial_l$, for $i,j,k,l=1,2,3$.}, one ends up with the
equation like that for charges in NCQED.

The result here by the Wong's approach in arbitrary background field
has been pointed out also via the behavior of open strings ending on D-branes in presence of B-field
\cite{jab-2}, and also obtained through the studying of implications of
possible noncommutativity in the present world in some specific examples \cite{jab-4,jab-5}. From the string
theory point of view the situation can be described as below. For example,
the mode expansion of open string coordinates ending on a D2-brane is
given by \cite{jab-2}:
\bea
&&X^0=x^0 + p^0 \tau + \sum_{n\neq 0} a^0_n \frac{e^{-in\tau}}{n} \cos n\sigma\nonumber\\
&&X^i=x^i+(p^i\tau - B^i_j p^j \sigma) +
\sum_{n\neq 0} \frac{e^{-in\tau}}{n} ( i a^i_n \cos n\sigma + B^i_j a^j \sin n\sigma),\;\;\;\;i=1,2\nonumber\\
&&X^b=x^b + p^b \tau + \sum_{n\neq 0} a^b_n \frac{e^{-in\tau}}{n} \cos n\sigma, \;\;\;\;b=3,\cdots,9
\eea
in which $B^i_j$ are components of the B-field background. Now we
see that for even the case in which the oscillations are suppressed,
the distance between the ends of open strings on the D2-brane is not zero, appearing to be
$\Delta^i= X^i(\sigma=0,\tau) - X^i(\sigma=\pi, \tau)= \pi B^i_j p^j$, by which we
expect a dipole behavior due to $\pm$ charges we assign to the ends of oriented
open strings \cite{9510135,jab-2}. This behavior of open strings
has been suggestive to formulate a theory for fields of dipoles rather
than for fields that their quanta are particles \cite{jab-3}.
The rule of multiplication of fields for dipoles is reminiscent of the $\star$-product.

\vspace{.3cm}
{\bf Acknowledgement:} The author is grateful to M. M. Sheikh-Jabbari
for a final reading of the manuscript.

\end{document}